\documentstyle[twocolumn,psfig,rotating]{mn}
\def\pmb#1{\setbox0=\hbox{#1}%
\kern-.025em\copy0\kern-\wd0
\kern.05em\copy0\kern-\wd0
\kern-.025em\raise.0433em\box0}

\begin{document}
\title[DM]
{New limits on the generation of magnetic field}
\author[Chuzhoy ]{Leonid Chuzhoy\\
Physics Department and The Space Research Institute,
Technion, Haifa 32000, Israel\\
E-mail: cleonid@tx.technion.ac.il
}

\maketitle

\begin{abstract}
Magnetic fields are generated in ionized objects rotating with respect
to a radiation background.  Based on conservation of canonical ion
momentum, it has been previously suggested that even if the radiation
intensity is unrestricted the maximum field strength generated this way
is $\sim 10^{-4}\Omega$ G, where $\Omega$ is the initial angular
velocity of the object. We show that this limit is valid only for an
object made of fully ionized and optically thin plasma.  The limit can
be relaxed by orders of magnitude in the presence of a high neutral
fraction or if the plasma is coupled to other forms of matter such as
stars or compact clouds.

\end{abstract}


\section {Introduction}
In a variety of astrophysical systems the observed strength of
magnetic field is close to an equipartition value.  Generally it is believed that the present values were
reached by dynamo amplification of an initially weak seed field (Widrow 2002).  However, the mechanism responsible for generating
the seed fields remains uncertain. Along with other possible
candidates a radiation drag has been proposed. In plasma moving with
respect to a radiation field the drag force acting on electrons is
stronger than on protons, causing the generation of electric currents. 
These currents 
 give rise to a magnetic field in the system. Mishustin \& Ruzmaikin
(1972) estimated that the value of magnetic fields generated in
protogalaxies by radiation drag is in the range $10^{-18} - 6\times
10^{-17}$ G. However, a subsequent analysis by Balbus(1993)
implied that, independent of the radiation intensity,  the
maximal value of the magnetic field  generated by radiation drag
is  $\sim m_{\rm p} c\Omega/e$, where $\Omega$ is the initial
angular velocity. For a typical protogalactic disk this limit is 
$\sim 10^{-19}$ G, which might be too small to account for the required seed field (Kulsrud 1999, but see Widrow 2002, Davis, Lilley \& Tornkvist 1999). According to Balbus this limit also invalidates theoretical estimates of magnetic field in several other systems, such as bipolar jets in protostellar systems (Chloe \& Jones 1993) and disks around massive compact objects (Walker 1988).

In this paper, however, we show that this limit is
valid only if the plasma is fully ionized, optically thin, and uncoupled
to any other form of matter. If on the other hand a large fraction of the
gas is neutral, condensed in optically thick clouds, or coupled to
another form of matter (stars, dark matter etc) then this limit might
have to be relaxed by several orders of magnitude.

\section{Equations}
\subsection{Two-component system}
Lets assume first that plasma is fully ionized, optically thin and
uncoupled to any other form of matter. For simplicity we assume that
plasma is composed only of protons and electrons.  In self-gravitating
astrophysical ionized objects global electric forces must be
similar in magnitude to the gravitational forces. This means that the
large-scale deviations from neutrality $(n_{\rm p}-n_{\rm e})/n_{\rm e}$ are close to 
$Gm_{\rm p}^2/e^2\sim 10^{-38}$, 
 the ratio between the electrical and gravitational forces. 
This is a tiny number and therefore  we assume that the plasma is 
locally neutral, i.e., $n_{\rm e}=n_{\rm p}$. 
 Similarly, because the magnetic field of an object of
size $R$ is proportional to $R(n_{\rm e}V_{\rm e}-n_{\rm p}V_{\rm p})$, for large objects the
requirement of a small magnetic field translates into nearly equal
velocities for electrons and protons, i.e., $V_{\rm e}=V_{\rm p}$. Also for large ionized objects we can assume that electrons and protons are tied together by induction, neglecting the collisional friction. 
 Using these
assumptions we write the equations of motions for protons and
electrons as follows:
\begin{eqnarray}
\label{acc}
\frac{dV_{\rm e}}{dt}&=&\frac{e}{m_{\rm p}}(E-\frac{B\times V_{\rm e}}{c})+\nabla\phi+\frac{\nabla (n_{\rm e} kT)}{n_{\rm e} m_{\rm p}},\\
\frac{dV_{\rm e}}{dt}&=&\frac{e}{m_{\rm e}}(\frac{B\times V_{\rm e}}{c}-E)+\nabla\phi+\frac{\nabla (n_{\rm e} kT)}{n_{\rm e} m_{\rm e}}+\frac{F_{\rm rad}}{m_{\rm e}},
\end{eqnarray}
where $\phi$ is the gravitational potential and $F_{\rm rad}$ is 
the radiation drag force on each electron.
 Combining both equations we find the electric field
\begin{equation}
E=\frac{B\times V_{\rm e}}{c}+\frac{m_{\rm p}-m_{\rm e}}{m_{\rm p}+m_{\rm e}}\frac{\nabla (n_{\rm e}kT)}{en_{\rm e}}+\frac{m_{\rm p}}{m_{\rm p}+m_{\rm e}}\frac{F_{\rm rad}}{e}.
\end{equation}
Taking curl of both sides of the above equation and using 
Faraday equation gives
\begin{eqnarray}
\label{Far}
\frac{\partial B}{c\partial t}=\frac{\nabla\times V_{\rm e}\times B}{c}+
\frac{m_{\rm p}-m_{\rm e}}{m_{\rm p}+m_{\rm e}}\nabla\times\frac{\nabla(n_{\rm e}kT)}{en_{\rm e}} \nonumber\\+
\frac{m_{\rm p}}{m_{\rm p}+m_{\rm e}}\frac{\nabla\times F_{\rm rad}}{e}.
\end{eqnarray}
The first and the second term on the right side of (\ref{Far}) are
respectively the dynamo amplification and the battery mechanism
(Biermann 1950). Since we are interested in generation of magnetic
field due to radiation drag only, we will ignore these
terms. Thus we remain with
\begin{equation}
\label{Far2}
\frac{\partial B}{c\partial t}=\frac{m_{\rm p}}{m_{\rm p}+m_{\rm e}}\frac{\nabla\times F_{\rm rad}}{e}.
\end{equation}
The radiation drag will act only as long as plasma rotates with
respect to photonic fluid, i.e. until it loses all its angular
momentum.If we ignore possible contraction or expansion of the object
during the process, then integrating (\ref{Far2}) in time gives
\begin{equation}
\label{Far3}
B_{\rm max}=\frac{m_{\rm p}c}{m_{\rm p}+m_{\rm e}}\frac{\nabla\times (P_e+P_p)}{e}=\frac{\nabla\times (m_{\rm p}c V_0)}{e},
\end{equation}
where $V_0$ and is the initial rotational velocity of plasma and $P_e$
and $P_p$ are the initial momenta of electrons and protons. Since
$\Omega=\nabla\times V$ is the angular velocity we get
\begin{equation}
\label{Far4}
{\rm B_{\rm max}=\frac{m_{\rm p} c\Omega_0 }{e}\approx 10^{-4}\Omega_0 \; G}
\end{equation}
Similar limits were  previously 
obtained by several authors (Harrison 1970; Balbus 1993).

\subsection{Three-component system}
 We now introduce a third component into the system, which interacts
with electrons and protons but not with radiation. 
The relevant equations of
motion are now
\begin{eqnarray}
\label{acc2}
m_{\rm p}\frac{dV_{\rm e}}{dt}&=&eE+K_{\rm p}(V_{\rm x}-V_{\rm e}),\\
\label{acc2a}
m_{\rm e}\frac{dV_{\rm e}}{dt}&=&-eE+K_{\rm e}(V_{\rm x}-V_{\rm e})+F_{\rm rad}, \\
(m_{\rm p}+m_{\rm e})\frac{dV_{\rm x}}{dt}&=&\chi(K_{\rm p}+K_{\rm e})(V_{\rm e}-V_{\rm x})\; ,
\end{eqnarray}
where $\chi$ is the mass density ratio 
of the  plasma to  the third component, $V_x$ is the velocity of the third component, and 
  $K_{\rm e}$ and $K_{\rm p}$ are, respectively,  coupling constants of the electrons and 
protons to the third component.  Combining the equations we obtain
\begin{equation}
\label{mult1}
eE\left(\frac{1}{m_{\rm e}}+\frac{1}{m_{\rm p}}\right)=
\frac{m_{\rm p}+m_{\rm e}}{\chi\left(K_{\rm e}+K_{\rm p}\right)}\left(\frac{K_{\rm p}}{m_{\rm p}}-
\frac{K_{\rm e}}{m_{\rm e}}\right)\frac{dV_{\rm x}}{dt}+\frac{F_{\rm rad}}{m_{\rm e}} \; .
\end{equation}
If there is a  strong coupling between the plasma and the third 
component, or equivalently the timescale of 
the drag $m_{\rm p} V_{\rm e}/F_{\rm rad}$ is much larger than $m_{\rm p}/(K_{\rm p}+K_{\rm e})$, then the
 velocity of the third component is close to the velocity of plasma, i.e.,
 $V_{\rm e}\sim V_{\rm x}$. Then 
\begin{equation}
\label{mult2}
(1+\chi^{-1})(m_{\rm p}+m_{\rm e})\frac{dV_{\rm e}}{dt}=F_{\rm rad}
\end{equation}
Combining eqs. \ref{mult1} and 
\ref{mult2}  we obtain
\begin{equation}
eE=\left(m_{\rm p}+\frac{K_{\rm p}}{K_{\rm p}+K_{\rm e}}\frac{m_{\rm p}+m_{\rm e}}{\chi}\right)\frac{dV_{\rm e}}{dt}.
\end{equation}
Using Faraday's law and integrating in time we derive the following 
upper limit on  the magnetic field 
\begin{equation}
\label{Far5}
B_{\rm max}=\left(1+\frac{(1+m_{\rm e}/m_{\rm p})K_{\rm p}}{\chi(K_{\rm e}+K_{\rm p})}\right)\frac{m_{\rm p} c \Omega_0}{e} \; .
\end{equation}
If  $K_{\rm p}\gg K_{\rm e}$, as is the case for coupling with neutral gas, 
then this limit reduces to 
\begin{equation}
\label{Far6}
B_{\rm max}\sim(1+\chi^{-1})\frac{m_{\rm p} c\Omega_0}{e} \; .
\end{equation}

\section{Conclusions}

The new limit is therefore larger by a factor of $(1+\chi^{-1})$. One
should not  conclude, however, that the magnetic field becomes unbounded
when $\chi$ goes to zero.  For an object of a fixed mass and angular
momentum decreasing $\chi$ is equivalent to decreasing $n_{\rm e}$, so for
too small $\chi$ induction becomes weaker than friction, which contradicts
 our initial assumption. In this case, according to eq. \ref{acc2a} 
the maximal velocity of the
electrons relative to the protons is  $F_{\rm rad}/K_{\rm e}$ and the
upper limit on magnetic field is obtained   from
\begin{equation}
\label{Far7}
\nabla\times B_{\rm max}=\frac{4\pi}{c}j=\frac{4\pi e}{c}n_{\rm e}\frac{F_{\rm rad}}{K_{\rm e}}
\end{equation}
For an object with a radius $R$ this yields
\begin{equation}
\label{Far8}
B_{\rm max}=\frac{4\pi e}{c}n_{\rm e}\frac{F_{\rm rad}}{K_{\rm e}}R
\end{equation}
Since $K_{\rm e}$ is proportional to the density of the third component then 
 $B_{\rm max}\propto n_{\rm e}/K_{\rm e} \propto \chi$.

As the upper limit on $B$ varies non-monotonically with $\chi$, the
maximal magnetic field is reached  for some intermediate value
of $\chi$, which  depends on the physical parameters of the system
(temperature, radiation strength, size of the object etc). Still in
large astrophysical systems $\chi$ needs to be very small before
eq.\ref{Far6} becomes invalid and thus the new limit would be much
larger than the previous one. Taking protogalaxies in the
post-recombination era as an example we find that since the fraction
of ionized gas is of order $10^{-4}$ the new upper limit is $\sim
10^{-15}{\rm G}$, which is consistent with Mishustin \& Ruzmaikin results.
\section*{acknowledgment}
I thank Adi Nusser for stimulating discussions. This research is supported
by a grant from the German Israeli Foundation for Scientific
Research and Development.

\end{document}